\documentclass[leqno]{article}

\PassOptionsToPackage{numbers, compress}{natbib}

    \usepackage[final]{ml4mols_2020}

\usepackage[utf8]{inputenc} 
\usepackage[T1]{fontenc}    
\usepackage{hyperref}       
\usepackage{url}            
\usepackage{booktabs}       
\usepackage{amsfonts}       
\usepackage{nicefrac}       
\usepackage{microtype}      
\usepackage{graphicx}
\usepackage{wrapfig}
\usepackage{multirow}
\usepackage{caption}
\usepackage{subcaption}
\usepackage{float}
\usepackage[table]{xcolor}

\usepackage{xcolor}

\usepackage{amsmath}
\usepackage{amsthm}

\newtheorem{proposition}{Proposition}

\usepackage{standalone}

\newcommand{\Real}{{\mathbb R}}

\title{Message Passing Networks for Molecules with Tetrahedral Chirality}

\author{
  Lagnajit Pattanaik\\
  Department of Chemical Engineering\\
  Massachusetts Institute of Technology\\
  Cambrdige, MA 02139 \\
  \texttt{lagnajit@mit.edu} \\
   \And
  Octavian-Eugen Ganea \\
  Computer Science and Artificial \\ Intelligence Laboratory\\
  Massachusetts Institute of Technology\\
  Cambridge, MA 02139 \\
  \texttt{oct@mit.edu} \\
   \And
  Ian Coley \\
  Department of Mathematics\\
  Rutgers University\\
  Piscataway, NJ 08854 \\
  \texttt{iacoley@math.rutgers.edu} \\
   \And
  Klavs F.~Jensen* \\
  Department of Chemical Engineering\\
  Massachusetts Institute of Technology\\
  Cambridge, MA 02139 \\
  \texttt{kfjensen@mit.edu} \\
   \And
  William H.~Green* \\
  Department of Chemical Engineering\\
  Massachusetts Institute of Technology\\
  Cambridge, MA 02139 \\
  \texttt{whgreen@mit.edu} \\
   \And
  Connor W.~Coley* \\
  Department of Chemical Engineering\\
  Massachusetts Institute of Technology\\
  Cambridge, MA 02139 \\
  \texttt{ccoley@mit.edu} \\
}

\begin{document}

\maketitle


\begin{abstract}
Molecules with identical graph connectivity can exhibit different physical and biological properties if they exhibit \textit{stereochemistry}--a spatial structural characteristic. However, modern neural architectures designed for learning structure-property relationships from molecular structures treat molecules as graph-structured data and therefore are invariant to stereochemistry. Here, we develop two custom aggregation functions for message passing neural networks to learn properties of molecules with tetrahedral chirality, one common form of stereochemistry. We evaluate performance on synthetic data as well as a newly-proposed protein-ligand docking dataset with relevance to drug discovery. Results show modest improvements over a baseline sum aggregator, highlighting opportunities for further architecture development.
\end{abstract}


\section{Introduction}

Recent advances in machine learning for chemistry that enable learning directly from molecular structures without preprocessing them into fixed-length vectors 
have yielded improvements in diverse settings including property prediction \cite{jimenez2020, muratov2020, chuang2020}, predictive synthesis \cite{coley_accounts}, and molecular optimization \cite{elton2019, brown2019}. However, stereochemistry has remained largely unexplored through the lens of deep learning despite being a significant aspect of molecular representation. 
Stereoisomers are molecular structures with the same graph connectivity but different spatial arrangements. 
Many different types of stereochemistry exist, ranging from common forms such as tetrahedral chirality and cis/trans isomerism to rarer forms such as atropisomerism and (the recently discovered) akamptisomerism \cite{canfield2018}. Each of them influences the accessible conformations of a small molecule, which in turn can influence the molecule's properties. Stereoisomers can have different physical properties (e.g., melting and boiling point), pharmacokinetics (e.g., absorption, distribution, metabolism, and excretion), and bioactivity (e.g., protein affinity) \cite{ariens1986, jamali1989, nguyen2006, chhabra2013}. 

This work focuses on tetrahedral chirality and its influence on property prediction. In principle, a molecule exhibiting tetrahedral chirality is neither a 2D graph nor a 3D structure, but somewhere in between. That is, the stereochemical designation of a chiral center limits the range of accessible conformers, but does not lock in a single 3D structure. Thus, we view using 3D representations to address tetrahedral chirality as too restrictive for conformationally-flexible molecules. 

Message passing neural networks (MPNNs) operate on molecules as graphs by treating atoms as nodes and bonds as edges \cite{duvenaud}. 
MPNNs operate by iteratively aggregating neighbor representations; traditional aggregation functions developed for graph-structured data such as sum, mean, and max are symmetric operators. Because stereoisomers have identical graph connectivities, symmetric aggregators operating on two different chiral centers will collapse their neighbors to an identical representation, regardless of chirality; 
that is, aggregation functions are the primary barrier to effectively realizing chirality in MPNN architectures. Equation \ref{eq:mpnn} shows a generic MPNN update with $x_i^{(k)} \in \mathbb{R}^F$ denoting the representation of node $i$ at layer $k$ and $x_j,~ j \in \mathcal{N}(i)$ denoting its neighbors. $\gamma$ amd $\phi$ represent differentiable functions while $e_{j,i} \in \mathbb{R}^F$ is the directed edge representation from node $j$ to $i$. $AGG$ refers to a generic aggregation scheme.


\begin{equation}\label{eq:mpnn}
    x_i^{(k)} = \gamma^{(k)} \left( x_i^{(k-1)}, x_i^{*(k)} \right), \quad x_i^{*(k)} = AGG_{j \in \mathcal{N}(i)} \phi^{(k)} (x_i^{(k-1)}, x_j^{(k-1)}, e_{j,i}^{(k-1)} )
\end{equation}


Here, we focus on extending MPNNs to remove the unwanted invariance to tetrahedral chirality. We develop and evaluate our methods on a new dataset that exemplifies the influence of tetrahedral chirality on molecular properties. Our primary contributions are: 
(1) Establishing a customized graph representation where edges of each node are ordered, not merely sets; 
(2) Proposing two theory-motivated methods to distinguish molecules with different tetrahedral chiralities through asymmetric node aggregation while still achieving physically-meaningful invariances;
(3) Proposing a benchmark dataset where the property of interest is a function of stereochemical configuration; and
(4) Evaluating empirical performance on this and other datasets using the proposed aggregation schemes through multiple criteria. We use the PyTorch Geometric framework to implement our network \cite{ptg} and make our code publicly available at \url{https://github.com/PattanaikL/chiral_gnn}.


\section{Related Work}

\begin{wrapfigure}{R}{0.3\textwidth}
    \vspace{-50pt}
    \centering
    \includegraphics[width=0.28\textwidth]{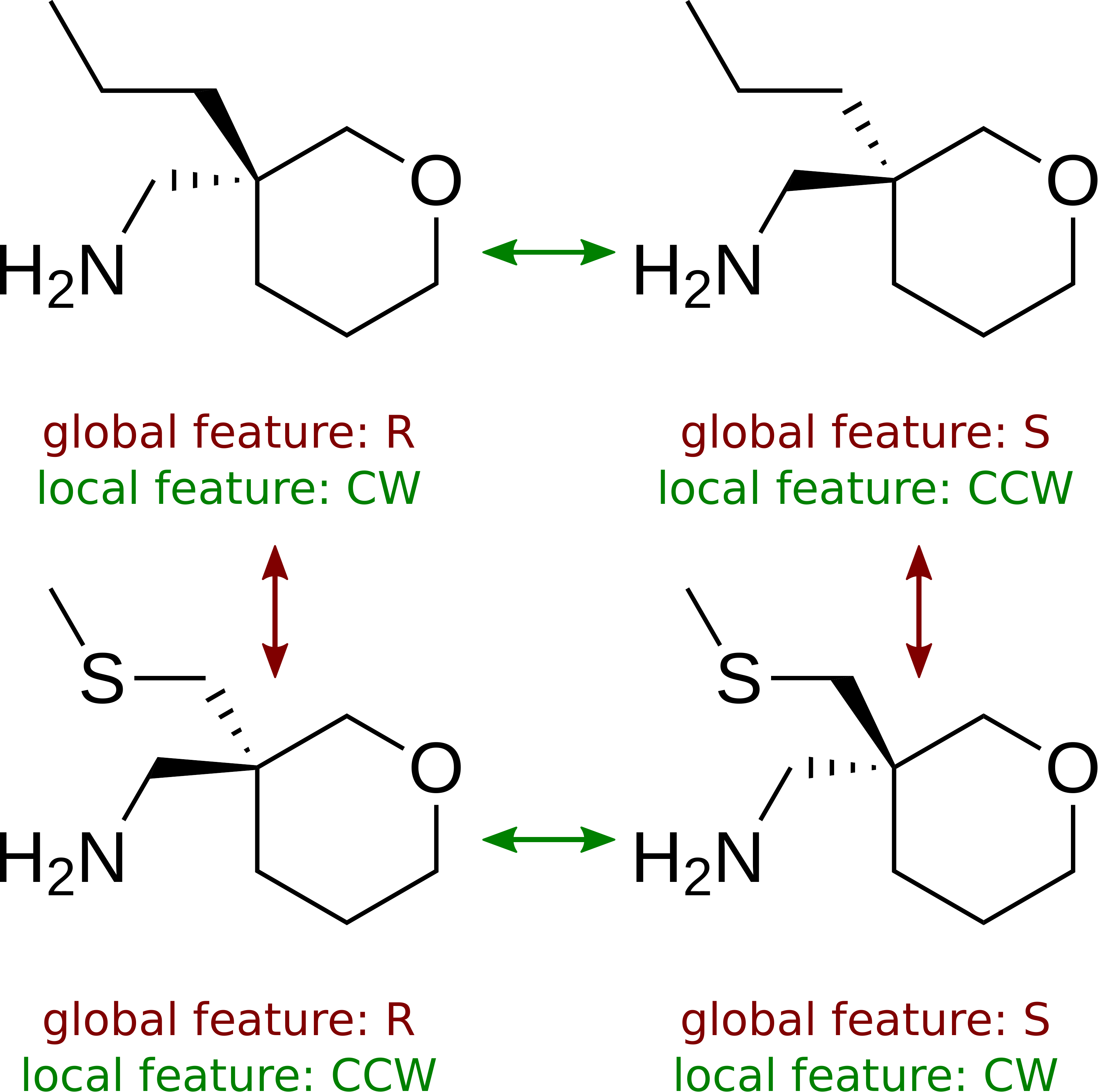}
    \caption{Global R/S features vs. local CW/CCW features}
    \label{fig:loc_v_glob}
    \vspace{-30pt}
\end{wrapfigure}

\paragraph{Message passing on 2D molecular graphs}
MPNNs have gained relevance for their application to molecular property prediction \cite{duvenaud, kearnes, gilmer} and are now a staple of many QSAR approaches \cite{gilmer2020}. Recent innovations include directional messages \cite{dmpnn} and hierarchical messages \cite{morris}. Many of these 2D graph methods do not address stereochemistry at all; in fact, the recent work by \citet{flam}, which includes 3D information, is one of the first MPNN architectures to explicitly address stereochemistry by performing message passing and aggregation over larger subgraphs.

\paragraph{Learning molecular invariances}
Achieving \{in,equi\}variances for molecules with neural networks has traditionally focused on rotational and translational symmetries, primarily for computed properties like electronic energy, which are identical between enantiomers. Architectures leading this effort include DTNNs \cite{dtnn}, SchNet \cite{schnet}, Comorant \cite{comorant}, and the recent DimeNet \cite{dimenet}. However, all of these networks operate in $\mathbb{R}^3$, which does not reflect the conformational flexibility of molecules.


\section{Approach}

\paragraph{Preliminaries}



Ostensibly, MPNNs might capture stereochemistry through atom- or bond-level descriptors, such as an R/S atom feature to distinguish enantiomers (i.e., non-superimposable stereoisomers that are mirror images of each other). These features represent \emph{global} measures of chirality, determined through CIP side-chain ranking conventions \cite{cip}. Figure \ref{fig:loc_v_glob} illustrates why global chirality descriptors as atom features are insufficient for MPNNs to meaningfully differentiate these structures.
\begin{wrapfigure}{r}{0.4\textwidth}
    \centering
    \includegraphics[width=0.38\textwidth]{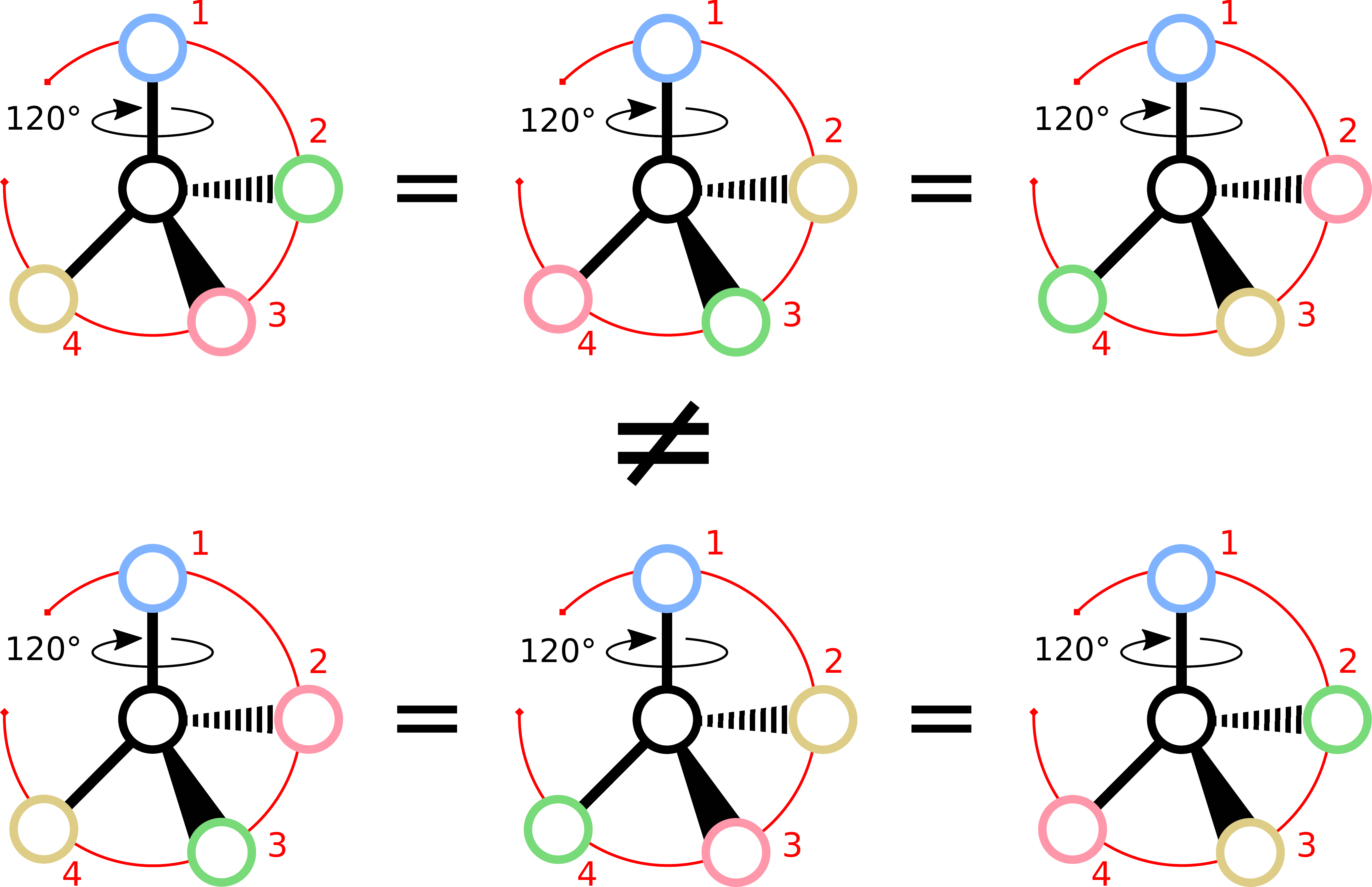}
    \caption{Explanation of how to categorize local chiral groups given an initial atom ordering. The top row represents one grouping of local chirality, while the bottom row represents another; rotations about all 4 bonds returns 12 permutations for each chiral group.}
    \label{fig:perms}
    \vspace{+10pt}
\end{wrapfigure}
With global descriptors, the two "R" structures would receive the same atom-level feature while the two "S" structures would receive a different feature. However, 
the local configurations at the tetrahedral center are different for each of the two “R” structures, despite sharing the same global descriptor. 

An alternate method captures the relevant information through \emph{local} chirality descriptors, denoted by clockwise/counter-clockwise labels (CW/CCW) in Figure~\ref{fig:loc_v_glob} as used in SMILES notation \cite{smiles}. Chiral centers receive a parity label (CW or CCW) based on the order of neighboring atoms, e.g. as provided in the SMILES string \cite{daylight}, relative to their spatial orientation, which assigns them to a local \emph{chiral group} of 12 equivalent permutations of four neighbors out of 24 total (Figure~\ref{fig:perms}). 
Directly using this parity bit as an atom feature would make the MPNN sensitive to arbitrary bookkeeping and SMILES canonicalization conventions. We therefore use this local parity bit in combination with the ordered set of neighbors to preserve physically-meaningful invariances. 


\paragraph{Theory}
We now prove the model structure necessary to constitute a universal approximator of all aggregation functions defined over the same chiral group. 
For any set $\mathcal{X}$, we define the quadruples set $\mathcal{X}^4 = \{ (x_1, x_2, x_3, x_4), x_i \in \mathcal{X}, i=1,2,3,4 \}$. We define the equivalence relation $\sim_C$ for a given chiral group $C$ as $(x_1, x_2, x_3, x_4) \sim_C (x'_1, x'_2, x'_3, x'_4)$ iff $\exists \pi \in \mathcal{P}(C)$ such that $(x_1, x_2, x_3, x_4) = (x'_{\pi(1)}, x'_{\pi(2)}, x'_{\pi(3)}, x'_{\pi(4)})$, where $\mathcal{P}(C)$ is the set of all 12 permutations of quadruples that are in the $C$ chiral group. Finally, we define the quotient set $\mathcal{C}_4(\mathcal{X}) = \mathcal{X}^4 / \sim_C$. 
Then, the following proposition proves the \textbf{universal approximation capability} of our proposed architecture. See the Appendix for a proof of the proposition.

\begin{proposition}
Any function $f : \mathcal{C}_4(\mathcal{X}) \rightarrow \Real^n$ is a valid function on quadruples from a chiral group C, i.e. it is invariant to any permutation from $\mathcal{P}(C)$, \textbf{iff} it can be decomposed in the form 
\begin{equation}
    f(x_1, x_2, x_3, x_4) = h\left(\sum_{\pi \in \mathcal{P}(C)} g( x_{\pi(1)}, x_{\pi(2)}, x_{\pi(3)}, x_{\pi(4)} ) \right)    
\label{eq:prop}
\end{equation}
for suitable functions $g$ and $h$.
\end{proposition}

\paragraph{Models}

Following Equation \ref{eq:prop}, we propose two novel aggregation functions for updating the representation of a tetrahedral center $x_i$ by adjusting the function $g$.  
Each  aggregation function allows enantiomers to be distinguished while remaining invariant to arbitrary bookkeeping conventions. 


\textit{Permutation (PERM)}. 
PERM 
introduces four separate weight matrices ($\{W_{g_i}\}_{i=1}^4$) for each ordered neighbor in a given permutation ($\{\pi_i\}_{i=1}^4$) and applies them to each of the 12 permutations of the chiral group (as denoted by $\pi \in \mathcal{P}(i)$). Note that in Equation \ref{eq:perm}, $\tau_{g_1}$ and $\tau_{g_2}$ represent different nonlinear operators (tanh and ReLU, respectively, in our case).

\begin{multline}\label{eq:perm}
    \small
    x_i^{*(k)} = W_{h_2} \tau_h \Biggl(\frac{1}{3} W_{h_1} \sum_{\pi \in \mathcal{P}(i)} \tau_{g_2} \Bigl( \tau_{g_1} (W_{g_1} x_{\pi(1)}^{(k-1)}) + \tau_{g_1} (W_{g_2} x_{\pi(2)}^{(k-1)}) + \tau_{g_1} (W_{g_3} x_{\pi(3)}^{(k-1)}) \\
    + \tau_{g_1} (W_{g_4} x_{\pi(4)}^{(k-1)}) \Bigr) \Biggr)
\end{multline}

\textit{Permutation-concatenation (PERM\_CAT)}. 
PERM\_CAT is similar to the PERM architecture, but rather than using four separate weight matrices, we use a single weight matrix and concatenate atom representations in each permutation. PERM\_CAT and PERM are equivalent when $\tau_{g_1}$ is linear. 

\begin{equation}\label{eq:permcat}
    x_i^{*(k)} = W_{h_2} \tau_h \left(\frac{1}{3} W_{h_1} \sum_{\pi \in \mathcal{P}(i)} \tau_g \Bigl( W_g (x_{\pi(1)}^{(k-1)} \| x_{\pi(2)}^{(k-1)} \| x_{\pi(3)}^{(k-1)} \| x_{\pi(4)}^{(k-1)}) \Bigr)\right)
\end{equation}


We test each aggregation function within several message passing frameworks to assess empirical performance. These include the standard graph convolutional network (GCN) \cite{gcn} modified to include edge features, the graph isomorphism network (GIN) \cite{gin} modified to use edge features \cite{gine} , and the directed message passing neural network (DMPNN) \cite{dmpnn}. We also test the effect of including only atom-level local chiral features describing CW/CCW, only global chiral features describing R/S, and including both sets of features; we denote these runs by ``LCF,'' ``GCF,'' and ``BCF,'' respectively. A third aggregation function that achieves the correct invariances but is not motivated by Proposition~1, the Product of Differences, can be found in the Appendix.


\section{Experiments}

\paragraph{Datasets}
Most property prediction datasets lack clean stereochemical information on which to benchmark chirality-aware models. 
To this end, we 
extract a subset of data from the D4 dopamine receptor protein-ligand docking screen by \citet{lyu}. Stereoisomers may exhibit distinct interaction energies when complexing with a protein of interest due to the conformations/poses they can achieve. Our dataset (D4DCHP) narrows the original 138 million molecules to stereoisomer pairs for a single 1,3-{\bf d}i{\bf c}yclo{\bf h}exyl{\bf p}ropane skeletal scaffold. We further define two additional subsets: one where enantiomers exhibit $>5$ kcal/mol differences in docking score (DIFF5) and one where molecules have a single tetrahedral center (CHIRAL1).  We also evaluate performance on the ``Lipo'' dataset of log-scaled partition coefficients from \citet{molnet} as 1127 of 4200 molecules in this dataset contain at least one tetrahedral center; unlike the D4DCHP dataset, however, we do not have complete pairs of enantiomers. See Appendix Section \ref{sec:lipo_results} for Lipo results. Additional dataset details can also be found in the Appendix.

\paragraph{Metrics}
We evaluate empirical performance of our custom aggregators on each dataset using root mean squared error (RMSE). However, as we are especially interested in how our methods separate stereoisomers in the D4DCHP dataset, we additionally evaluate the model's ability to correctly order two enantiomers; 
predictions within 0.1 kcal/mol of each other are considered indistinguishable and treated as a random guess with 50\% accuracy. 
We denote this task \emph{rank classification}.

\begin{wraptable}{r}{.35\textwidth}
    \setlength{\tabcolsep}{3pt}
    \vspace{-12pt}
    \small
    \caption{Classification of R/S stereocenters of CHIRAL1 test set (\%)}\label{tab:rs}
    \begin{tabular}{lccc}
        \toprule
        \multirow{2}[3]{*}{\parbox{1cm}{Aggregation method}} & \multicolumn{3}{c}{Graph architecture} \\
        \cmidrule(r){2-4}
             & GCN & GIN & DMPNN \\
        \midrule
        SUM         & 50            & 50            & 50  \\
        PERM        & \textbf{100}  & \textbf{100}  & \textbf{98}  \\
        PERM\_CAT   & \textbf{100}  & \textbf{100}  & \textbf{98}  \\
        \bottomrule
    \end{tabular}
    \vspace{5pt}
\end{wraptable}

\paragraph{R/S Classification}

To ensure that our new aggregation methods understand tetrahedral chirality, we first evaluate them on the simple task of classifying molecules from CHIRAL1 as R or S. 
This tests if the MPNNs can (1) distinguish enantiomers and (2) learn CIP rules to assign an R/S label.  
Because the input atom/bonds features and graph connectivity are identical for a pair of stereoisomers, we expect a sum aggregator to achieve an accuracy of 50\%. Empirical results (Table \ref{tab:rs}) support this hypothesis. Sum aggregators across MPNN architectures do not perform better than a random classifier, while both custom aggregators for all three architectures reach nearly perfect accuracy when classifying molecules as R or S. 

\paragraph{D4DCHP}

We next evaluate empirical performance on the full D4DCHP dataset (Table \ref{tab:rmse_rc}). We shuffle the data randomly but ensure that stereoisomer pairs are kept together in the train/validation/test splits.  In the absence of atom-level chiral features, the asymmetric aggregation methods of PERM and PERM\_CAT provide a significant benefit over the SUM aggregator baseline. 

The performance of the different aggregation methods in the presence of chiral features depends on the architecture. The custom aggregation functions show measurable improvements in both rank classification and RMSE for the GCN architecture, but these improvements are not so clear for the GIN and DMPNN architectures.  For the GIN, including both atom-level chiral features is sufficient to capture the trends picked up by the custom aggregators in the absence of such features. The custom aggregators yield improved performance for the DMPNN in terms of rank classification but not RMSE. 

The strong influence of atom-level chiral features is rather surprising. By including just two additional atom-level features, the sum aggregators achieve approximately a 10\% improvement in rank classification, regardless of graph architecture. The more expressive graph architectures can more effectively use this chiral information, such that the explicit equivariances offered by the custom aggregators no longer provide a significant benefit. Regardless, the highest observed rank classification accuracy of  62\% provides clear room for improvement, motivating further research into representation learning methods that can better address tetrahedral chirality.

\begin{table}[H]
  \vspace{-10pt}
  \setlength{\tabcolsep}{3pt}
  \small
  \caption{Rank classification percent for test set (trained on full D4DCHP, results for DIFF5 subset) and full D4DCHP RMSE for test set}
  \label{tab:rmse_rc}
  \centering
  \begin{tabular}{lcccccc}
    \toprule
    \multirow{2}[3]{*}{\parbox{1cm}{Aggregation method}} & \multicolumn{3}{c}{DIFF5 rank classification accuracy (\%)} & \multicolumn{3}{c}{D4DCHP error (RMSE)} \\
    \cmidrule(r){2-4} \cmidrule(r){5-7}
         & GCN & GIN & DMPNN  & GCN & GIN & DMPNN  \\
    \midrule
    SUM             & 50.0 $\pm$ 0.0            & 50.0 $\pm$ 0.0            & 50.0 $\pm$ 0.0           & 6.77 $\pm$ 0.05            & 6.45 $\pm$ 0.07           & 6.45 $\pm$ 0.06   \\
    SUM (LCF)       & 55.6 $\pm$ 0.1            & 57.9 $\pm$ 0.3            & 57.8 $\pm$ 0.3           & 6.76 $\pm$ 0.04            & 6.42 $\pm$ 0.06           & 6.41 $\pm$ 0.06   \\
    SUM (GCF)       & 57.9 $\pm$ 0.5            & 61.5 $\pm$ 0.2            & 60.6 $\pm$ 0.0           & 6.73 $\pm$ 0.05            & 6.38 $\pm$ 0.06           & 6.39 $\pm$ 0.06   \\
    SUM (BCF)       & 58.7 $\pm$ 0.7            & 62.1 $\pm$ 0.1            & 60.9 $\pm$ 0.4           & 6.95 $\pm$ 0.30            & 6.38 $\pm$ 0.06           & 6.37 $\pm$ 0.06   \\
    \arrayrulecolor{black!30}\midrule
    PERM            & 58.3 $\pm$ 0.5            & 58.6 $\pm$ 0.3            & 59.7 $\pm$ 0.5           & 6.69 $\pm$ 0.06            & 6.44 $\pm$ 0.06           & 6.42 $\pm$ 0.05   \\
    PERM (LCF)      & 60.4 $\pm$ 0.5            & 61.1 $\pm$ 0.1            & 61.5 $\pm$ 0.1           & 6.69 $\pm$ 0.07            & 6.41 $\pm$ 0.06           & 6.41 $\pm$ 0.05   \\
    PERM (GCF)      & 59.8 $\pm$ 0.5            & 61.2 $\pm$ 0.4            & 61.5 $\pm$ 0.3           & 6.68 $\pm$ 0.06            & 6.40 $\pm$ 0.07           & 6.40 $\pm$ 0.05   \\
    PERM (BCF)      & 60.7 $\pm$ 0.3            & 61.5 $\pm$ 0.2            & 61.3 $\pm$ 0.2           & 6.67 $\pm$ 0.06            & 6.39 $\pm$ 0.06           & 6.39 $\pm$ 0.06   \\
    \arrayrulecolor{black!30}\midrule
    PERM\_CAT       & 60.5 $\pm$ 0.2            & 60.5 $\pm$ 0.3            & 61.2 $\pm$ 0.3           & 6.65 $\pm$ 0.06            & 6.40 $\pm$ 0.05           & 6.40 $\pm$ 0.06   \\
    PERM\_CAT (LCF) & 61.0 $\pm$ 0.1            & 62.0 $\pm$ 0.3            & 61.3 $\pm$ 0.3           & 6.62 $\pm$ 0.06            & 6.38 $\pm$ 0.06           & 6.40 $\pm$ 0.05   \\
    PERM\_CAT (GCF) & 60.7 $\pm$ 0.2            & 61.8 $\pm$ 0.5            & 61.8 $\pm$ 0.2           & 6.64 $\pm$ 0.05            & 6.40 $\pm$ 0.07           & 6.38 $\pm$ 0.06   \\
    PERM\_CAT (BCF) & 60.8 $\pm$ 0.1            & 62.2 $\pm$ 0.1            & 62.4 $\pm$ 0.2           & 6.65 $\pm$ 0.06            & 6.37 $\pm$ 0.07           & 6.39 $\pm$ 0.07   \\
    \arrayrulecolor{black}\bottomrule
  \end{tabular}
  \vspace{-10pt}
\end{table}


\section{Conclusion and Outlook}


We have developed two aggregation functions to learn properties for molecules with tetrahedral chirality, motivated by a general expression for a universal function approximator. The custom aggregators can fully distinguish enantiomers on a toy R/S classification problem and show comparable or modestly-improved performance over a baseline sum aggregator on a newly-proposed D4DHCP dataset depending on the MPNN architecture and inclusion of atom-level chiral features.

Empirical performance is a function of both (a) whether the network understand stereochemistry and (b) the extent to which stereochemistry matters for that dataset and property. Many existing datasets are inadequate; either stereochemistry is not relevant (e.g., QM9 energies \cite{qm9}), or the data are not rich enough to convey the nuanced trends yielded by chirality (e.g., lipophilicity). Thus, until such datasets become available, we recommend that algorithm development occur in controlled environments like D4DCHP. Finally, we have only attempted to address tetrahedral chirality for MPNNs here; we leave stereoisomerism beyond tetrahedral chirality to future work.



\begin{ack}
We gratefully acknowledge the Machine Learning for Pharmaceutical Discovery and Synthesis consortium for funding this work. We also acknowledge the MIT SuperCloud and Lincoln Laboratory Supercomputing Center for providing HPC resources that have contributed to the research results reported within this paper \cite{supercloud}.
\end{ack}

\bibliographystyle{plainnat}
\bibliography{neurips_2020}

\begin{thebibliography}{36}
\providecommand{\natexlab}[1]{#1}
\providecommand{\url}[1]{\texttt{#1}}
\expandafter\ifx\csname urlstyle\endcsname\relax
  \providecommand{\doi}[1]{doi: #1}\else
  \providecommand{\doi}{doi: \begingroup \urlstyle{rm}\Url}\fi

\bibitem[Akiba et~al.(2019)Akiba, Sano, Yanase, Ohta, and Koyama]{optuna}
Takuya Akiba, Shotaro Sano, Toshihiko Yanase, Takeru Ohta, and Masanori Koyama.
\newblock Optuna: A next-generation hyperparameter optimization framework.
\newblock In \emph{Proceedings of the 25rd {ACM} {SIGKDD} International
  Conference on Knowledge Discovery and Data Mining}, 2019.

\bibitem[Anderson et~al.(2019)Anderson, Hy, and Kondor]{comorant}
Brandon Anderson, Truong~Son Hy, and Risi Kondor.
\newblock Cormorant: Covariant molecular neural networks.
\newblock In \emph{Advances in Neural Information Processing Systems}, pages
  14537--14546, 2019.

\bibitem[Ariens(1986)]{ariens1986}
EJ~Ariens.
\newblock Chirality in bioactive agents and its pitfalls.
\newblock \emph{Trends in Pharmacological Sciences}, 7:\penalty0 200--205,
  1986.

\bibitem[Brown et~al.(2019)Brown, Fiscato, Segler, and Vaucher]{brown2019}
Nathan Brown, Marco Fiscato, Marwin~HS Segler, and Alain~C Vaucher.
\newblock Guacamol: benchmarking models for de novo molecular design.
\newblock \emph{Journal of chemical information and modeling}, 59\penalty0
  (3):\penalty0 1096--1108, 2019.

\bibitem[Canfield et~al.(2018)Canfield, Blake, Cai, Luck, Krausz, Kobayashi,
  Reimers, and Crossley]{canfield2018}
Peter~J Canfield, Iain~M Blake, Zheng-Li Cai, Ian~J Luck, Elmars Krausz, Rika
  Kobayashi, Jeffrey~R Reimers, and Maxwell~J Crossley.
\newblock A new fundamental type of conformational isomerism.
\newblock \emph{Nature chemistry}, 10\penalty0 (6):\penalty0 615--624, 2018.

\bibitem[Chhabra et~al.(2013)Chhabra, Aseri, and Padmanabhan]{chhabra2013}
Naveen Chhabra, Madan~L Aseri, and Deepak Padmanabhan.
\newblock A review of drug isomerism and its significance.
\newblock \emph{International journal of applied and basic medical research},
  3\penalty0 (1):\penalty0 16, 2013.

\bibitem[Chuang et~al.(2020)Chuang, Gunsalus, and Keiser]{chuang2020}
Kangway~V Chuang, Laura Gunsalus, and Michael~J Keiser.
\newblock Learning molecular representations for medicinal chemistry.
\newblock \emph{Journal of Medicinal Chemistry}, 2020.

\bibitem[Coley et~al.(2018)Coley, Green, and Jensen]{coley_accounts}
Connor~W Coley, William~H Green, and Klavs~F Jensen.
\newblock Machine learning in computer-aided synthesis planning.
\newblock \emph{Accounts of chemical research}, 51\penalty0 (5):\penalty0
  1281--1289, 2018.

\bibitem[Cross and Klyne(2013)]{cip}
LC~Cross and W~Klyne.
\newblock \emph{Rules for the nomenclature of Organic Chemistry: Section E:
  stereochemistry (Recommendations 1974)}.
\newblock Elsevier, 2013.

\bibitem[Daylight Chemical Information~Systems(2011)]{daylight}
Inc. Daylight Chemical Information~Systems.
\newblock Smiles - a simplified chemical language, 2011.
\newblock URL
  \url{https://www.daylight.com/dayhtml/doc/theory/theory.smiles.html}.

\bibitem[Duvenaud et~al.(2015)Duvenaud, Maclaurin, Iparraguirre, Bombarell,
  Hirzel, Aspuru-Guzik, and Adams]{duvenaud}
David~K Duvenaud, Dougal Maclaurin, Jorge Iparraguirre, Rafael Bombarell,
  Timothy Hirzel, Al{\'a}n Aspuru-Guzik, and Ryan~P Adams.
\newblock Convolutional networks on graphs for learning molecular fingerprints.
\newblock In \emph{Advances in neural information processing systems}, pages
  2224--2232, 2015.

\bibitem[Elton et~al.(2019)Elton, Boukouvalas, Fuge, and Chung]{elton2019}
Daniel~C Elton, Zois Boukouvalas, Mark~D Fuge, and Peter~W Chung.
\newblock Deep learning for molecular design—a review of the state of the
  art.
\newblock \emph{Molecular Systems Design \& Engineering}, 4\penalty0
  (4):\penalty0 828--849, 2019.

\bibitem[Fey and Lenssen(2019)]{ptg}
Matthias Fey and Jan~Eric Lenssen.
\newblock Fast graph representation learning with pytorch geometric.
\newblock \emph{arXiv preprint arXiv:1903.02428}, 2019.

\bibitem[Flam-Shepherd et~al.(2020)Flam-Shepherd, Wu, Friederich, and
  Aspuru-Guzik]{flam}
Daniel Flam-Shepherd, Tony Wu, Pascal Friederich, and Alan Aspuru-Guzik.
\newblock Neural message passing on high order paths.
\newblock \emph{arXiv preprint arXiv:2002.10413}, 2020.

\bibitem[Gilmer et~al.(2017)Gilmer, Schoenholz, Riley, Vinyals, and
  Dahl]{gilmer}
Justin Gilmer, Samuel~S Schoenholz, Patrick~F Riley, Oriol Vinyals, and
  George~E Dahl.
\newblock Neural message passing for quantum chemistry.
\newblock \emph{arXiv preprint arXiv:1704.01212}, 2017.

\bibitem[Gilmer et~al.(2020)Gilmer, Schoenholz, Riley, Vinyals, and
  Dahl]{gilmer2020}
Justin Gilmer, Samuel~S Schoenholz, Patrick~F Riley, Oriol Vinyals, and
  George~E Dahl.
\newblock Message passing neural networks.
\newblock In \emph{Machine Learning Meets Quantum Physics}, pages 199--214.
  Springer, 2020.

\bibitem[Hu et~al.(2019)Hu, Liu, Gomes, Zitnik, Liang, Pande, and
  Leskovec]{gine}
Weihua Hu, Bowen Liu, Joseph Gomes, Marinka Zitnik, Percy Liang, Vijay Pande,
  and Jure Leskovec.
\newblock Strategies for pre-training graph neural networks.
\newblock \emph{arXiv preprint arXiv:1905.12265}, 2019.

\bibitem[Jamali et~al.(1989)Jamali, Mehvar, and Pasutto]{jamali1989}
F~Jamali, R~Mehvar, and FM~Pasutto.
\newblock Enantioselective aspects of drug action and disposition: therapeutic
  pitfalls.
\newblock \emph{Journal of pharmaceutical sciences}, 78\penalty0 (9):\penalty0
  695--715, 1989.

\bibitem[Jim{\'e}nez-Luna et~al.(2020)Jim{\'e}nez-Luna, Grisoni, and
  Schneider]{jimenez2020}
Jos{\'e} Jim{\'e}nez-Luna, Francesca Grisoni, and Gisbert Schneider.
\newblock Drug discovery with explainable artificial intelligence.
\newblock \emph{arXiv preprint arXiv:2007.00523}, 2020.

\bibitem[Kearnes et~al.(2016)Kearnes, McCloskey, Berndl, Pande, and
  Riley]{kearnes}
Steven Kearnes, Kevin McCloskey, Marc Berndl, Vijay Pande, and Patrick Riley.
\newblock Molecular graph convolutions: moving beyond fingerprints.
\newblock \emph{Journal of computer-aided molecular design}, 30\penalty0
  (8):\penalty0 595--608, 2016.

\bibitem[Kipf and Welling(2016)]{gcn}
Thomas~N Kipf and Max Welling.
\newblock Semi-supervised classification with graph convolutional networks.
\newblock \emph{arXiv preprint arXiv:1609.02907}, 2016.

\bibitem[Klicpera et~al.(2020)Klicpera, Gro{\ss}, and G{\"u}nnemann]{dimenet}
Johannes Klicpera, Janek Gro{\ss}, and Stephan G{\"u}nnemann.
\newblock Directional message passing for molecular graphs.
\newblock \emph{arXiv preprint arXiv:2003.03123}, 2020.

\bibitem[Landrum et~al.(2006)]{rdkit}
Greg Landrum et~al.
\newblock Rdkit: Open-source cheminformatics.
\newblock 2006.

\bibitem[Lyu et~al.(2019)Lyu, Wang, Balius, Singh, Levit, Moroz, O’Meara,
  Che, Algaa, Tolmachova, et~al.]{lyu}
Jiankun Lyu, Sheng Wang, Trent~E Balius, Isha Singh, Anat Levit, Yurii~S Moroz,
  Matthew~J O’Meara, Tao Che, Enkhjargal Algaa, Kateryna Tolmachova, et~al.
\newblock Ultra-large library docking for discovering new chemotypes.
\newblock \emph{Nature}, 566\penalty0 (7743):\penalty0 224--229, 2019.

\bibitem[Morris et~al.(2019)Morris, Ritzert, Fey, Hamilton, Lenssen, Rattan,
  and Grohe]{morris}
Christopher Morris, Martin Ritzert, Matthias Fey, William~L Hamilton, Jan~Eric
  Lenssen, Gaurav Rattan, and Martin Grohe.
\newblock Weisfeiler and leman go neural: Higher-order graph neural networks.
\newblock In \emph{Proceedings of the AAAI Conference on Artificial
  Intelligence}, volume~33, pages 4602--4609, 2019.

\bibitem[Muratov et~al.(2020)Muratov, Bajorath, Sheridan, Tetko, Filimonov,
  Poroikov, Oprea, Baskin, Varnek, Roitberg, et~al.]{muratov2020}
Eugene~N Muratov, J{\"u}rgen Bajorath, Robert~P Sheridan, Igor~V Tetko, Dmitry
  Filimonov, Vladimir Poroikov, Tudor~I Oprea, Igor~I Baskin, Alexandre Varnek,
  Adrian Roitberg, et~al.
\newblock Qsar without borders.
\newblock \emph{Chemical Society Reviews}, 2020.

\bibitem[Nguyen et~al.(2006)Nguyen, He, and Pham-Huy]{nguyen2006}
Lien~Ai Nguyen, Hua He, and Chuong Pham-Huy.
\newblock Chiral drugs: an overview.
\newblock \emph{International journal of biomedical science: IJBS}, 2\penalty0
  (2):\penalty0 85, 2006.

\bibitem[Ramakrishnan et~al.(2014)Ramakrishnan, Dral, Rupp, and
  Von~Lilienfeld]{qm9}
Raghunathan Ramakrishnan, Pavlo~O Dral, Matthias Rupp, and O~Anatole
  Von~Lilienfeld.
\newblock Quantum chemistry structures and properties of 134 kilo molecules.
\newblock \emph{Scientific data}, 1\penalty0 (1):\penalty0 1--7, 2014.

\bibitem[Reuther et~al.(2018)Reuther, Kepner, Byun, Samsi, Arcand, Bestor,
  Bergeron, Gadepally, Houle, Hubbell, et~al.]{supercloud}
Albert Reuther, Jeremy Kepner, Chansup Byun, Siddharth Samsi, William Arcand,
  David Bestor, Bill Bergeron, Vijay Gadepally, Michael Houle, Matthew Hubbell,
  et~al.
\newblock Interactive supercomputing on 40,000 cores for machine learning and
  data analysis.
\newblock In \emph{2018 IEEE High Performance extreme Computing Conference
  (HPEC)}, pages 1--6. IEEE, 2018.

\bibitem[Sch{\"u}tt et~al.(2017{\natexlab{a}})Sch{\"u}tt, Kindermans, Felix,
  Chmiela, Tkatchenko, and M{\"u}ller]{schnet}
Kristof Sch{\"u}tt, Pieter-Jan Kindermans, Huziel Enoc~Sauceda Felix, Stefan
  Chmiela, Alexandre Tkatchenko, and Klaus-Robert M{\"u}ller.
\newblock Schnet: A continuous-filter convolutional neural network for modeling
  quantum interactions.
\newblock In \emph{Advances in neural information processing systems}, pages
  991--1001, 2017{\natexlab{a}}.

\bibitem[Sch{\"u}tt et~al.(2017{\natexlab{b}})Sch{\"u}tt, Arbabzadah, Chmiela,
  M{\"u}ller, and Tkatchenko]{dtnn}
Kristof~T Sch{\"u}tt, Farhad Arbabzadah, Stefan Chmiela, Klaus~R M{\"u}ller,
  and Alexandre Tkatchenko.
\newblock Quantum-chemical insights from deep tensor neural networks.
\newblock \emph{Nature communications}, 8\penalty0 (1):\penalty0 1--8,
  2017{\natexlab{b}}.

\bibitem[Weininger(1988)]{smiles}
David Weininger.
\newblock Smiles, a chemical language and information system. 1. introduction
  to methodology and encoding rules.
\newblock \emph{Journal of chemical information and computer sciences},
  28\penalty0 (1):\penalty0 31--36, 1988.

\bibitem[Wu et~al.(2018)Wu, Ramsundar, Feinberg, Gomes, Geniesse, Pappu,
  Leswing, and Pande]{molnet}
Zhenqin Wu, Bharath Ramsundar, Evan~N Feinberg, Joseph Gomes, Caleb Geniesse,
  Aneesh~S Pappu, Karl Leswing, and Vijay Pande.
\newblock Moleculenet: a benchmark for molecular machine learning.
\newblock \emph{Chemical science}, 9\penalty0 (2):\penalty0 513--530, 2018.

\bibitem[Xu et~al.(2018)Xu, Hu, Leskovec, and Jegelka]{gin}
Keyulu Xu, Weihua Hu, Jure Leskovec, and Stefanie Jegelka.
\newblock How powerful are graph neural networks?
\newblock \emph{arXiv preprint arXiv:1810.00826}, 2018.

\bibitem[Yang et~al.(2019)Yang, Swanson, Jin, Coley, Eiden, Gao, Guzman-Perez,
  Hopper, Kelley, Mathea, et~al.]{dmpnn}
Kevin Yang, Kyle Swanson, Wengong Jin, Connor Coley, Philipp Eiden, Hua Gao,
  Angel Guzman-Perez, Timothy Hopper, Brian Kelley, Miriam Mathea, et~al.
\newblock Analyzing learned molecular representations for property prediction.
\newblock \emph{Journal of chemical information and modeling}, 59\penalty0
  (8):\penalty0 3370--3388, 2019.

\bibitem[Zaheer et~al.(2017)Zaheer, Kottur, Ravanbakhsh, Poczos, Salakhutdinov,
  and Smola]{zaheer2017deep}
Manzil Zaheer, Satwik Kottur, Siamak Ravanbakhsh, Barnabas Poczos, Russ~R
  Salakhutdinov, and Alexander~J Smola.
\newblock Deep sets.
\newblock In \emph{Advances in neural information processing systems}, pages
  3391--3401, 2017.

\end{thebibliography}

\newpage

\section{Appendix}

\subsection{Theory}
\subsubsection{Proof of proposition 1}
\textit{Proof:}
First, we easily observe that the functional form in Eq.~\ref{eq:prop} is a valid function on $\mathcal{C}_4(\mathcal{X})$.

Conversely, we note that there is a bijection between $\mathcal{C}_4(\mathcal{X})$ and the set of sets of C-chiral group permutations of a quadruple. This bijection is given by the following function:

\begin{equation*}
\phi(x_1, x_2, x_3, x_4) = \{ (x_{\pi(1)}, x_{\pi(2)}, x_{\pi(3)},  x_{\pi(4)}) \mid \pi \in \mathcal{P}(C) \}
\end{equation*}

Using \cite{zaheer2017deep}, we know that all functions from sets $Y \subset \mathcal{Y}$ to $\Real^n$ have the form $h(\sum_{y \in Y} \psi(y))$ for some functions $h$ and $\psi$. Taking $\mathcal{Y} = \mathcal{X}^4$ and $g = \psi \circ \phi$, we obtain the desired result. \qed

\subsection{Models}
\subsubsection{Product of differences (PD)}

Separate from the PERM and PERM\_CAT aggregators, we develop an additional aggregator that achieves the desired equivariances between local chiral groups without using multiple permutations. This method takes pairwise differences between neighbor atom representations and then computes the signed geometric mean. Note that the inversion of the tetrahedral center (e.g., by exchanging two neighbors) will invert the sign of the resulting product. 

\begin{alignat*}{2}\label{eq:pd}
    x_i^{*(k)} & = \tau_h && \left(W_h \prod_{j \in \mathcal{N}(i)} \prod_{m \in \mathcal{N}(i), m < j} ( x_j^{(k-1)} - x_m^{(k-1)} ) ^{1/6}\right) \\
    & = \tau_h && \Bigl( W_h (x_1^{(k-1)} - x_2^{(k-1)}) ^{1/6} \times (x_1^{(k-1)} - x_3^{(k-1)}) ^{1/6} \\
    & &&\times (x_1^{(k-1)} - x_4^{(k-1)}) ^{1/6} \times (x_2^{(k-1)} - x_3^{(k-1)}) ^{1/6} \\
    & &&\times (x_2^{(k-1)} - x_4^{(k-1)}) ^{1/6} \times (x_3^{(k-1)} - x_4^{(k-1)}) ^{1/6} \Bigr)
\end{alignat*}

\subsubsection{Aggregation visualization}

\begin{figure}[H]
    \centering
    \includegraphics[width=0.8\textwidth]{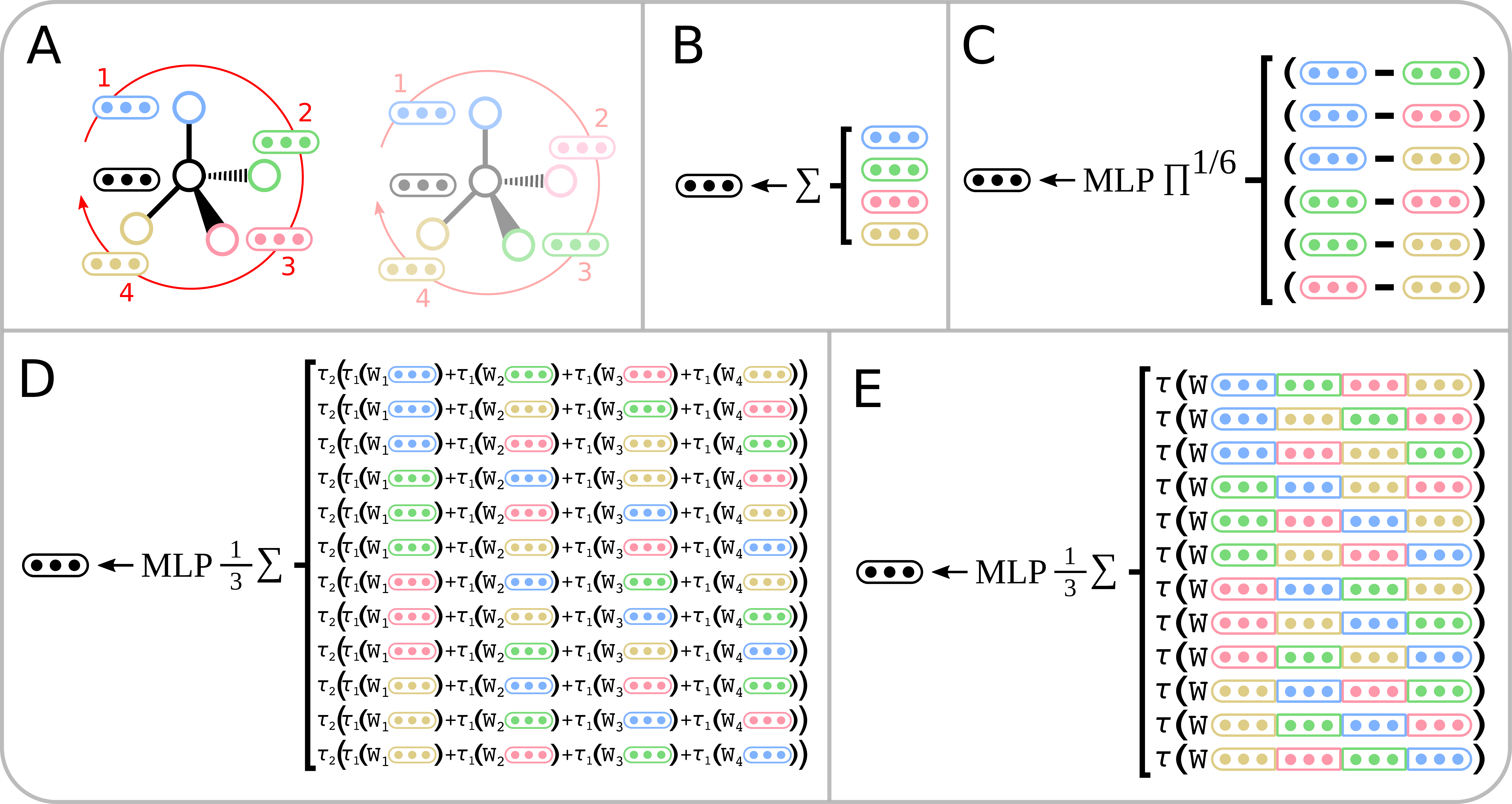}
    \caption{A. Two stereoisomers and their atom feature vectors  representing different local chiral groups. B-E. SUM, PD, PERM, and PERM\_CAT aggregation examples for first chiral group.}
    \label{fig:message_passing}
\end{figure}

Figure \ref{fig:message_passing} shows an example of the sum aggregator and all custom aggregators applied to a tetrahedral chiral center for a given initial ordering.

\subsubsection{Chiral group permutations}

Table \ref{tab:all_perms} shows all permutations for both chiral groups. Any initial atom ordering falls into one of these two groups, which determines all permutations considered for the chiral message. Note that, in practice, we adjust all atom orderings to fall within a single group (Group 1 in this case), so we only need to apply a single set of MLP functions (corresponding to $g$ and $h$ in Equation \ref{eq:prop}). This adjustment only changes the bookkeeping used for tetrahedral centers, not the physical meaning. See Figure \ref{fig:perms} for a visual explanation on how to obtain these permutation groups.

\begin{table}[H]
    \centering
    \caption{Chiral group permutations}\label{tab:all_perms}
    \begin{tabular}{cc}
        \toprule
        Group 1 & Group 2\\
        \midrule
        \{0, 1, 2, 3\}, \{0, 2, 3, 1\}, \{0, 3, 1, 2\} & \{0, 1, 3, 2\}, \{0, 2, 1, 3\}, \{0, 3, 2, 1\} \\
        \{1, 0, 3, 2\}, \{1, 3, 2, 0\}, \{1, 2, 0, 3\} & \{1, 0, 2, 3\}, \{1, 3, 0, 2\}, \{1, 2, 3, 0\} \\
        \{2, 0, 1, 3\}, \{2, 1, 3, 0\}, \{2, 3, 0, 1\} & \{2, 0, 3, 1\}, \{2, 1, 0, 3\}, \{2, 3, 1, 0\} \\
        \{3, 0, 2, 1\}, \{3, 2, 1, 0\}, \{3, 1, 0, 2\} & \{3, 0, 1, 2\}, \{3, 2, 0, 1\}, \{3, 1, 2, 0\}\\
        \bottomrule
    \end{tabular}
\end{table}

\subsection{Datasets}

\subsubsection{D4DCHP dataset justification}
Evaluating chiral interactions on an enzyme binding pocket in theory is a complex task. While some stereoisomers will interact differently with the enzyme, others will show few differences. That is, the chiral portion of the small molecule structure may not be relevant to the binding thermodynamics, such that two enantiomers may even interact with identical binding energies. This uncertainty of whether or not chirality is relevant presents an important challenge, and one that must be present when designing a benchmark for this task.

\subsubsection{D4DCHP dataset construction}
We filter the D4 docking study from \cite{lyu} by the generalized Bemis-Murcko scaffold 1,3-{\bf d}i{\bf c}yclo{\bf h}exyl{\bf p}ropane (hence the dataset name D4DCHP). Additionally, we mandate that all molecules have at least one tetrahedral chiral center (i.e. we remove molecules with chiral centers with fewer than four neighbors\footnote{Chiral carbons with three heavy atom neighbors and one hydrogen are retained.}, such as with sulfoxides) and further filter the dataset to include only stereosiomer pairs (i.e., enantiomers or diastereomers), removing structures without such a complement in the dataset. The constitutes the full dataset. We also create two subsets of this full dataset: one retains pairs of enantiomers with differences in docking score greater than five kcal/mol (DIFF5) while the other retains molecules with only one tetrahedral center (CHIRAL1). Note that the provided \texttt{d4\_docking.csv} file contains the full dataset, while the \texttt{\{subset\}/split\{x\}.npy} files define the subsets.

\subsubsection{Synthetic R/S dataset generation}
For the toy R/S classification task, we use the CHIRAL1 dataset and label each pair of enantiomers with a 0/1 denoting that the single stereocenter is R/S. The assignment is made using RDKit version 2020.03.2 \cite{rdkit}. We keep pairs of stereoisomers together in the training/validation/testing sets but determine each split randomly, and we only run one split. The dataset is provided as \texttt{d4\_docking\_rs.csv} and the split we used for the results is \texttt{rs/split0.npy}.

\subsubsection{D4DCHP dataset statistics}

\begin{figure}[H]
    \centering
    \includegraphics[width=1.0\textwidth]{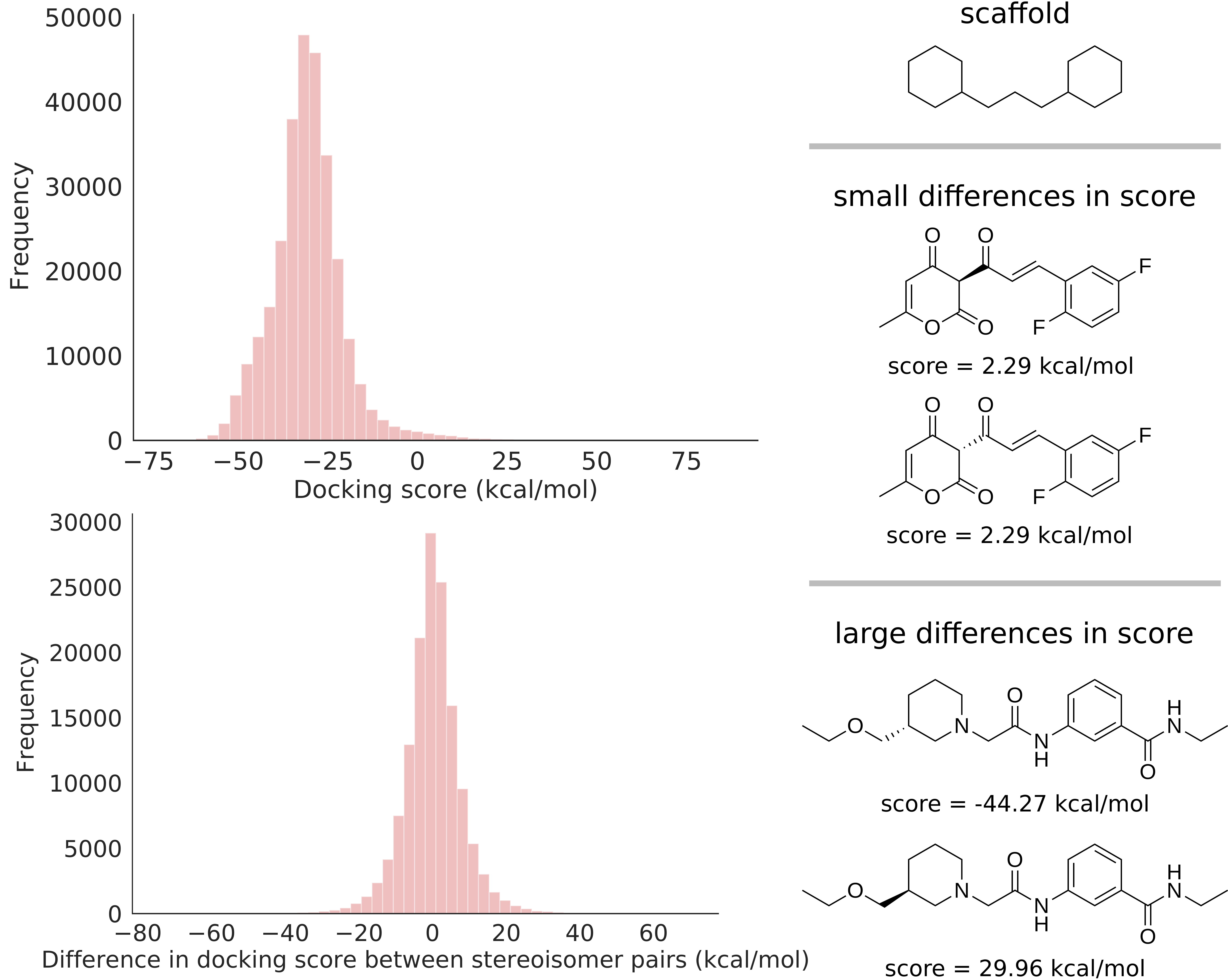}
    \caption{(Top left) Distribution of docking scores for full dataset. (Bottom left) Distribution of difference in docking score between pairs of stereosiomers. (Right) Generalized Bemis-Murcko scaffold used to subset data from original study and example structures with small and large differences in docking score.}
    \label{fig:dataset_stats}
\end{figure}

Table \ref{tab:stats} shows various statistics for the full dataset and its subsets. The top left of Figure \ref{fig:dataset_stats} shows the data distribution for the full set, highlighting the large range of values present. The bottom left of Figure \ref{fig:dataset_stats} shows the distribution obtained by taking the difference between all pairs on stereoisomers in the full dataset. This reinforces the earlier justification of the D4DCHP dataset; while some pairs show little to no relative difference in docking score, others show large differences. Thus, tetrahedral stereochemistry has a complex effect on this function, rendering this dataset a challenging but realistic one for development of chiral representation models. The rest of Figure \ref{fig:dataset_stats} shows the scaffold used to extract the data and example structures from D4DCHP.


\begin{table}[H]
    \centering
    \caption{D4DCHP dataset statistics}\label{tab:stats}
    \begin{tabular}{lccc}
        \toprule
        \multirow{2}[3]{*}{Statistics} & \multicolumn{3}{c}{Data subset} \\
        \cmidrule(r){2-4}
                & FULL & DIFF5 & CHIRAL1 \\
        \midrule
        Number datapoints               & 287468    & 119166    & 204778    \\
        Min (kcal/mol)                  & -71.29    & -71.29    & -67.54    \\
        Max (kcal/mol)                  & 86.94     & 86.94     & 86.94     \\
        Mean (kcal/mol)                 & -30.67    & -29.32    & -31.19    \\
        Standard deviation (kcal/mol)   & 10.07     & 11.86     & 9.91      \\
        \bottomrule
    \end{tabular}
\end{table}

\subsection{Experimental details}

\subsubsection{Featurization}
We convert each SMILES in the csv files to PyTorch Geometric data objects with node and edge features, detailed in Table \ref{tab:feats}, using RDKit \cite{rdkit}. For structures with multiple fragments, we only retain the largest fragment. This only affects one entry in the Lipo dataset and does not affect the D4DCHP dataset, as all molecules in the dataset consist of a single structure. We include only heavy atoms (i.e. non-hydrogen atoms) in the molecular graph. However, for chiral centers that include a hydrogen atom as one of the four substituents, we make the attached hydrogen explicit; this is the case for a significant number of structures in the dataset. This construction facilitates use of the custom aggregation functions, but note that even for the sum aggregator, we make such hydrogens explicit in the molecular graph.

\begin{table}[H]
    \centering
    \caption{Atom and bond features}\label{tab:feats}
    \begin{tabular}{cccc}
        \toprule
        \multicolumn{4}{c}{\multirow{2}{*}{Atom features}} \\ \\
        Indices & Description & Options & Type \\
        \midrule
        0-11    & atom identity                 & H, C, N, O, F, Si, P, S, Cl, Br, I, other         & one-hot  \\
        12-18   & degree                        & 0, 1, 2, 3, 4, 5, other                           & one-hot  \\
        19-24   & formal charge                 & -2, -1, 0, 1, 2, other                            & one-hot  \\
        25-30   & total number of Hs on atom    & 0, 1, 2, 3, 4, other                              & one-hot  \\
        31-36   & hybridization                 & $sp$, $sp^2$, $sp^3$, $sp^3d$, $sp^3d^2$, other   & one-hot  \\
        37      & aromaticity                   & true, false                                       & one-hot  \\
        38      & atomic mass                   & $\mathbb{R}_{>0}$                                 & value * 0.1  \\
        39-41   & local chirality (optional)    & CW, CCW, other                                    & one-hot  \\
        42-44   & global chirality (optional)   & R, S, other                                       & one-hot  \\
        \midrule
        \multicolumn{4}{c}{\multirow{2}{*}{Bond features}} \\ \\
        Indices & Description & Options & Type \\
        \midrule
        0    & bond presence                    & true, false                       & one-hot  \\
        1-4  & bond type                        & single, double, triple, aromatic  & one-hot  \\
        5    & conjugation                      & true, false                       & one-hot  \\
        6    & presence in ring                 & true, false                       & one-hot  \\
        \bottomrule
    \end{tabular}
\end{table}

\subsubsection{Tetrahedral centers}
We take advantage of RDKit's internal bookkeeping to assign CW/CCW labels to chiral tetrahedral centers \cite{rdkit}. Before calculating the updated node representation for a custom aggregation function, we artificially change all tetrahedral centers to the same chiral grouping so they can be described by the same parity bit. That is, for all tetrahedral centers marked as CCW, we switch the order of the first and second neighbors as defined by RDKit before applying permutations. This allows us to feed the all tetrahedral centers through consistently-parametrized matrices and ensures that the correct permutations are applied to all tetrahedral chiral centers.

\subsubsection{Training}
We split the data into training/validation/testing sets using an 80/10/10 split. The reported statistics are averages of three runs, each using different splits, and the sample standard deviation. We run each method for a fixed 80 epochs and retain the model with the best validation performance. For the the D4DCHP dataset, we keep pairs of stereoisomers together in the training/validation/testing sets but determine each split randomly. All models use the same seed to initialize network weights, ensuring consistency between runs.

We use the Noam scheduler implementation from \citet{dmpnn}, which increases the learning rate linearly to the max rate for two warm-up epochs, and decreases it exponentially afterwards to the final rate. In our implementation, we ask the user to supply the maximum rate, setting the initial and final rates to be a tenth of the max rate.

For the D4DCHP dataset, we choose hyperparameters by using a simple grid scan for the DMPNN graph architecture with the sum aggregator. We use the final hyperaparameters, which provide good performances across all base architectures, for all test runs with the other aggregators (Table \ref{tab:hyperparams}).

\begin{table}
    \centering
    \caption{D4DCHP hyerparameter choices}\label{tab:hyperparams}
    \begin{tabular}{lcc}
        \toprule
        Hyperparameter & Sample range & Final choice \\
        \midrule
        Hidden size         & 300, 600, 900, 1200       & 300   \\
        Depth               & 2, 3, 4, 5, 6             & 3     \\
        Dropout             & 0, 0.2, 0.4, 0.6, 0.8, 1  & 0.2   \\
        Max learning rate   & 1e-5, 1e-4, 1e-3          & 1e-4  \\
        Batch size          & 25, 50, 100               & 50    \\
        \bottomrule
    \end{tabular}
\end{table}

\subsection{Additional results}

\subsubsection{R/S Classification}

Table \ref{tab:rs_full} includes the R/S classification results for the product of differences aggregator, which shows that it too can perfectly distinguish between chiral centers.

\begin{table}
    \centering
    \caption{Classification of R/S stereocenters of CHIRAL1 test set for all custom aggregators (\%)}
    \label{tab:rs_full}
    \begin{tabular}{lccc}
        \toprule
        \multirow{2}[3]{*}{\parbox{1cm}{Aggregation method}} & \multicolumn{3}{c}{Graph architecture} \\
        \cmidrule(r){2-4}
             & GCN & GIN & DMPNN \\
        \midrule
        SUM         & 50            & 50            & 50  \\
        PD          & \textbf{100}  & \textbf{100}  & \textbf{100}  \\
        PERM        & \textbf{100}  & \textbf{100}  & 98  \\
        PERM\_CAT   & \textbf{100}  & \textbf{100}  & 98  \\
        \bottomrule
    \end{tabular}
\end{table}

\subsubsection{D4DCHP}

Table \ref{tab:full_results} shows the full validation and test RMSE values from the D4DCHP dataset. The validation results follow the same trends as the test results in that GCN architecture benefits from the custom aggregators, the DMPNN architecture shows improvements in rank classification but not RMSE with the custom aggregators, and the GIN architecture achieves good performance with only a sum aggregator and chiral features. RMSE results of the PD aggregation are consistently worse than others. While the method is constructed to achieve specific equivariances, a product of differences has little physical motivation relative to other aggregation schemes more akin to belief propagation. 

\begin{table}
  \small
  \caption{D4DCHP RMSE validation/test results for full dataset and rank classification results for DIFF5 subset}
  \label{tab:full_results}
  \centering
  \begin{tabular}{lccc}
    \toprule
    \multicolumn{4}{c}{\multirow{2}{*}{RMSE (validation/test)}} \\ \\
    \multirow{2}[3]{*}{\parbox{1cm}{Aggregation method}} & \multicolumn{3}{c}{Graph architecture} \\
    \cmidrule(r){2-4}
         & GCN & GIN & DMPNN \\
    \midrule
    SUM             & 6.75 $\pm$ 0.05 / 6.77 $\pm$ 0.05           & 6.43 $\pm$ 0.04 / 6.45 $\pm$ 0.07           & 6.41 $\pm$ 0.02 / 6.45 $\pm$ 0.06           \\
    SUM (LCF)       & 6.74 $\pm$ 0.06 / 6.76 $\pm$ 0.04           & 6.39 $\pm$ 0.03 / 6.42 $\pm$ 0.06           & 6.39 $\pm$ 0.02 / 6.41 $\pm$ 0.06           \\
    SUM (GCF)       & 6.72 $\pm$ 0.05 / 6.73 $\pm$ 0.05           & 6.36 $\pm$ 0.03 / 6.38 $\pm$ 0.06           & 6.36 $\pm$ 0.03 / 6.39 $\pm$ 0.06           \\
    SUM (BCF)       & 6.72 $\pm$ 0.07 / 6.95 $\pm$ 0.30           & 6.35 $\pm$ 0.04 / 6.38 $\pm$ 0.06           & 6.35 $\pm$ 0.03 / 6.37 $\pm$ 0.06           \\
    \arrayrulecolor{black!30}\midrule
    PD              & 7.02 $\pm$ 0.06 / 7.03 $\pm$ 0.05           & 6.62 $\pm$ 0.06 / 6.63 $\pm$ 0.04           & 6.64 $\pm$ 0.07 / 6.64 $\pm$ 0.06           \\
    PD (LCF)        & 7.02 $\pm$ 0.06 / 7.04 $\pm$ 0.06           & 6.64 $\pm$ 0.04 / 6.66 $\pm$ 0.05           & 6.64 $\pm$ 0.06 / 6.64 $\pm$ 0.06           \\
    PD (GCF)        & 7.00 $\pm$ 0.04 / 7.03 $\pm$ 0.05           & 6.64 $\pm$ 0.05 / 6.66 $\pm$ 0.05           & 6.59 $\pm$ 0.04 / 6.58 $\pm$ 0.05           \\
    PD (BCF)        & 7.01 $\pm$ 0.05 / 7.02 $\pm$ 0.05           & 6.65 $\pm$ 0.06 / 6.66 $\pm$ 0.06           & 6.57 $\pm$ 0.04 / 6.59 $\pm$ 0.05           \\
    \arrayrulecolor{black!30}\midrule
    PERM            & 6.67 $\pm$ 0.05 / 6.69 $\pm$ 0.06           & 6.41 $\pm$ 0.04 / 6.44 $\pm$ 0.06           & 6.40 $\pm$ 0.04 / 6.42 $\pm$ 0.05           \\
    PERM (LCF)      & 6.65 $\pm$ 0.05 / 6.69 $\pm$ 0.07           & 6.38 $\pm$ 0.04 / 6.41 $\pm$ 0.06           & 6.39 $\pm$ 0.04 / 6.41 $\pm$ 0.05           \\
    PERM (GCF)      & 6.67 $\pm$ 0.03 / 6.68 $\pm$ 0.06           & 6.38 $\pm$ 0.04 / 6.40 $\pm$ 0.07           & 6.37 $\pm$ 0.04 / 6.40 $\pm$ 0.05           \\
    PERM (BCF)      & 6.64 $\pm$ 0.04 / 6.67 $\pm$ 0.06           & 6.37 $\pm$ 0.04 / 6.39 $\pm$ 0.06           & 6.37 $\pm$ 0.04 / 6.39 $\pm$ 0.06           \\
    \arrayrulecolor{black!30}\midrule
    PERM\_CAT       & 6.61 $\pm$ 0.03 / 6.65 $\pm$ 0.06           & 6.37 $\pm$ 0.03 / 6.40 $\pm$ 0.05           & 6.38 $\pm$ 0.03 / 6.40 $\pm$ 0.06           \\
    PERM\_CAT (LCF) & 6.59 $\pm$ 0.04 / 6.62 $\pm$ 0.06           & 6.36 $\pm$ 0.04 / 6.38 $\pm$ 0.06           & 6.37 $\pm$ 0.03 / 6.40 $\pm$ 0.05           \\
    PERM\_CAT (GCF) & 6.61 $\pm$ 0.03 / 6.64 $\pm$ 0.05           & 6.36 $\pm$ 0.03 / 6.40 $\pm$ 0.06           & 6.36 $\pm$ 0.04 / 6.38 $\pm$ 0.06           \\
    PERM\_CAT (BCF) & 6.63 $\pm$ 0.04 / 6.65 $\pm$ 0.06           & 6.35 $\pm$ 0.03 / 6.37 $\pm$ 0.07           & 6.35 $\pm$ 0.03 / 6.39 $\pm$ 0.07           \\
    \arrayrulecolor{black}\midrule
    \multicolumn{4}{c}{\multirow{2}{*}{Rank classification (\%)}} \\ \\
    \multirow{2}[3]{*}{\parbox{1cm}{Aggregation method}} & \multicolumn{3}{c}{Graph architecture} \\
    \arrayrulecolor{black}\cmidrule(r){2-4}
         & GCN & GIN & DMPNN \\
    \arrayrulecolor{black}\midrule
    SUM             & 50.0 $\pm$ 0.0           & 50.0 $\pm$ 0.0           & 50.0 $\pm$ 0.0           \\
    SUM (LCF)       & 55.6 $\pm$ 0.1           & 57.9 $\pm$ 0.3           & 57.8 $\pm$ 0.3           \\
    SUM (GCF)       & 57.9 $\pm$ 0.5           & 61.5 $\pm$ 0.2           & 60.6 $\pm$ 0.0           \\
    SUM (BCF)       & 58.7 $\pm$ 0.7           & 62.1 $\pm$ 0.1           & 60.9 $\pm$ 0.4           \\
    \arrayrulecolor{black!30}\midrule
    PD              & 57.3 $\pm$ 0.9           & 58.9 $\pm$ 0.4           & 58.7 $\pm$ 0.2           \\
    PD (LCF)        & 57.6 $\pm$ 0.4           & 59.0 $\pm$ 0.4           & 58.8 $\pm$ 0.5           \\
    PD (GCF)        & 57.5 $\pm$ 0.5           & 58.2 $\pm$ 0.3           & 59.4 $\pm$ 0.6           \\
    PD (BCF)        & 58.1 $\pm$ 0.4           & 58.9 $\pm$ 0.2           & 58.7 $\pm$ 0.3           \\
    \arrayrulecolor{black!30}\midrule
    PERM            & 58.3 $\pm$ 0.5           & 58.6 $\pm$ 0.3           & 59.7 $\pm$ 0.5           \\
    PERM (LCF)      & 60.4 $\pm$ 0.5           & 61.1 $\pm$ 0.1           & 61.5 $\pm$ 0.1           \\
    PERM (GCF)      & 59.8 $\pm$ 0.5           & 61.2 $\pm$ 0.4           & 61.5 $\pm$ 0.3           \\
    PERM (BCF)      & 60.7 $\pm$ 0.3           & 61.5 $\pm$ 0.2           & 61.3 $\pm$ 0.2           \\
    \arrayrulecolor{black!30}\midrule
    PERM\_CAT       & 60.5 $\pm$ 0.2           & 60.5 $\pm$ 0.3           & 61.2 $\pm$ 0.3           \\
    PERM\_CAT (LCF) & 61.0 $\pm$ 0.1           & 62.0 $\pm$ 0.3           & 61.3 $\pm$ 0.3           \\
    PERM\_CAT (GCF) & 60.7 $\pm$ 0.2           & 61.8 $\pm$ 0.5           & 61.8 $\pm$ 0.2           \\
    PERM\_CAT (BCF) & 60.8 $\pm$ 0.1           & 62.2 $\pm$ 0.1           & 62.4 $\pm$ 0.2           \\
    \arrayrulecolor{black}\bottomrule
  \end{tabular}
\end{table}

\begin{figure}
    \centering
    \includegraphics[width=\textwidth]{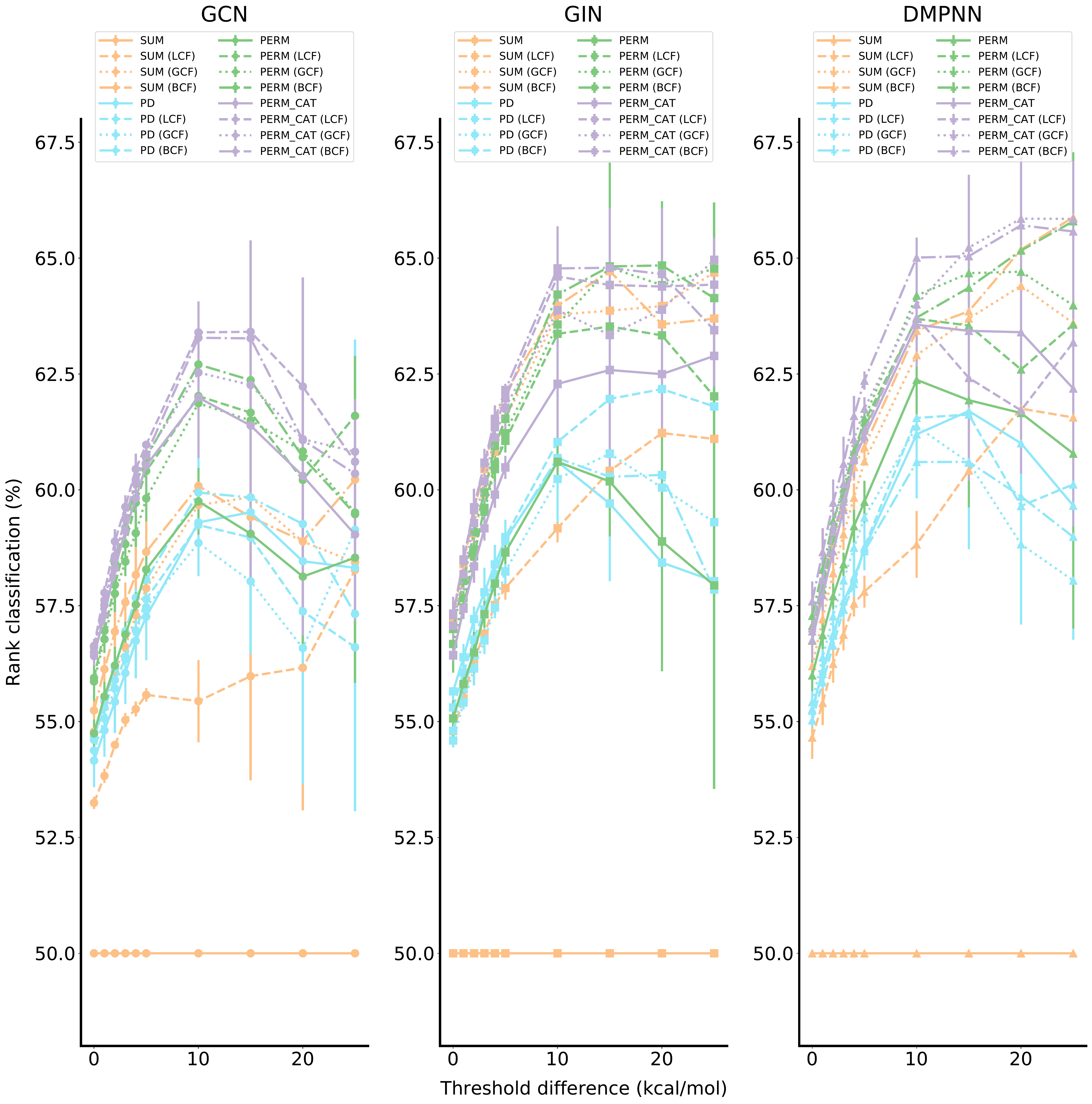}
    \caption{Rank classification accuracy for different test subsets selected based on a minimum threshold for the difference in ground truth docking score. Accuracy generally increases with larger ground truth differences between stereoisomers.}
    \label{fig:rank_thresh}
\end{figure}

Figure \ref{fig:rank_thresh} shows rank classification results for different threshold values (such that a threshold of 5~kcal/mol corresponds to the results in Table \ref{tab:rmse_rc}). The PERM\_CAT aggregator is consistently the best across graph architectures. With large differences in stereoisomers, all aggregators tend to increase rank classification accuracy for the GIN and DMPNN architectures. While the PD aggregator showed worse RMSE than the baseline SUM aggregator, it is able to rank stereoisomers correctly more than 50\% of the time.

\subsubsection{Lipophilicity}
\label{sec:lipo_results}

We also test our custom aggregation functions on the standard lipophilicity MoleculeNet dataset without pairs of stereoisomers present \cite{molnet}. Over 25\% of these data contain a chiral center, but because the dataset is so small and stereoisomer pairs are not present, we posit that the data are not rich enough for our aggregators to learn information about chirality. Table \ref{tab:lipo_rmse} shows these results, which agree with our hypothesis. Without a substantial number of chiral examples, the custom aggregators cannot learn nuanced trends and achieve similar performance to the baseline SUM aggregator. Note that the small size of the dataset allows us to perform a full hyperparameter optimization for each individual run, using the hyperparameters and sample ranges in Table \ref{tab:hyperparams} and a budget of 128 combinations. We perform this search using the Optuna framework \cite{optuna}.

\begin{table}[H]
  \small
  \caption{Lipo test RMSE}
  \label{tab:lipo_rmse}
  \centering
  \begin{tabular}{lccc}
    \toprule
    \multicolumn{4}{c}{\multirow{2}{*}{RMSE (validation/test)}} \\ \\
    \multirow{2}[3]{*}{\parbox{1cm}{Aggregation method}} & \multicolumn{3}{c}{Graph architecture} \\
    \cmidrule(r){2-4}
         & GCN & GIN & DMPNN \\
    \midrule
    SUM             & \textbf{0.63 $\pm$ 0.04}  & \textbf{0.64 $\pm$ 0.02}  & \textbf{0.60 $\pm$ 0.01}  \\
    SUM (LCF)        & 0.64 $\pm$ 0.03           & \textbf{0.64 $\pm$ 0.01}  & \textbf{0.60 $\pm$ 0.01}  \\
    PD              & 0.65 $\pm$ 0.01           & 0.67 $\pm$ 0.02           & 0.68 $\pm$ 0.02           \\
    PD (LCF)         & 0.67 $\pm$ 0.05           & 0.69 $\pm$ 0.03           & 0.67 $\pm$ 0.01           \\
    PERM            & 0.65 $\pm$ 0.05           & 0.66 $\pm$ 0.01           & 0.62 $\pm$ 0.01           \\
    PERM (LCF)       & 0.65 $\pm$ 0.02           & 0.65 $\pm$ 0.02           & 0.62 $\pm$ 0.01           \\
    PERM\_CAT       & 0.66 $\pm$ 0.03           & 0.65 $\pm$ 0.03           & 0.62 $\pm$ 0.01           \\
    PERM\_CAT (LCF)  & 0.67 $\pm$ 0.04           & 0.65 $\pm$ 0.01           & 0.62 $\pm$ 0.02           \\
    \bottomrule
  \end{tabular}
\end{table}

\end{document}